# Monitoring Student Activity in Collaborative Software Development

Emerging Patterns of Collaboration and Self-Organization

Daniel Dietsch, Andreas Podelski
Albert-Ludwigs-Universität Freiburg
Freiburg, Germany
{dietsch, podelski}@informatik.uni-freiburg.de

Jaechang Nam, Pantelis M. Papadopoulos, Martin Schäf
United Nations University
S.A.R Macau, China
{jcnam, pmpapad, schaef}@iist.unu.edu

*Abstract*—This paper presents data analysis from a course on Software Engineering in an effort to identify metrics and techniques that would allow instructor to act proactively and identify patterns of low engagement and inefficient peer collaboration. Over the last two terms, 106 students in their second year of studies formed 20 groups and worked collaboratively to develop video games. Throughout the lab, students have to use a variety of tools for managing and developing their projects, such as software version control, static analysis tools, wikis, mailing lists, etc. The students are also supported by weekly meetings with teaching assistants and instructors regarding group progress, code quality, and management issues. Through these meetings and their interactions with the software tools, students leave a detailed trace of data related to their individual engagement and their collaboration behavior in their groups. The paper provides discussion on the different sources of data that can be monitored, and present preliminary results on how these data can be used to analyze students' activity.

*Index Terms*—Project-based learning, monitoring, computer-supported collaborative learning, software development, video game development.

## I. Introduction

Software Engineering is a very demanding domain posing considerable complexity for both the students and the instructors. First, learning in the domain requires acquisition of deep domain knowledge and development of multifaceted skills, and their application in ill-structured settings. In other words, Software Engineering is highly contextualized and strongly anchored to the real world. As researchers on ill-structured domains suggest, the cognitive skills necessary for solving a problem in these domains are different from those required in well-structured domains [1], including extensive problem representation, active construction of the problem space, justification, monitoring and evaluating [2],[3],[4],[5].

Second, in the professional world, software development is rarely considered an individual activity. Software engineers typically work in groups playing different roles and adding different parts towards a problem solution. Similarly, students in Software Engineering need to understand and experience the social aspects of software development. The role of peer interaction has been emphasized over the years as most of the computer-supported collaborative learning (CSCL) literature draws from the socio-cultural approaches of learning, such as social constructivism [6],[7] and activity theory [8],[9]. Collaborative learning is focused on the social aspects of learning and the interactions that can affect the learning outcome [10]. CSCL has increasingly drawn attention as a teaching/learning approach and is frequently implemented at all levels of education [11]. Students have to learn how to collaborate with others, meaning that they need to be able to engage in meaningful interactions that would promote learning.

Third, the volume of practical and technical knowledge of the domain could be overwhelming for the students. Usually, a Software Engineering course assignment engages the students in playing different roles (e.g., requirements engineers, designers, analysts, coders, etc.). Through these roles, students have to learn how to use a range of tools and methods. Although learning the specifications of a tool may not be important for understanding the domain, it may nevertheless increase the students' workload.

The most common approach used by instructors to ease students in the intricacies of the Software Engineering domain is the project-based instruction. The literature abounds with variations of this approach. For example, Kilamo, Hammouda, and Chatti [12] incorporated a reputation system to support the social aspect of learning and enhance students' collaboration. Stamelos and his colleagues [13],[14] suggested the use of real open-source projects instead of the fictitious ones offered in the safe environment of the class. Čavrak, Orlić, and Crnković [15] applied the project-based approach in global software development courses for students in a distributed environment. MacKellar [16],[17] investigated the communication patterns of students in a class-wide project assignment. Van der Duim et al. [37] report their experience on international student projects with industry partners. Smarkusky et al. [35] propose self- and peer-assessment to identify "poor leaders" and "free riders" in project-based software engineering education. Hazzan and Dubinsky [36] propose pair programming sessions with students and supervisors as a solution to the free rider problem.

Over the last two years, sophomore students majoring in Computer Science at University of Freiburg has been engaged in group collaboration on a software development project-based learning activity, in which students are supported by instructors, teaching assistants (TAs), and a comprehensive list of software tools that help them monitor, organize, and manage their activity. Our broader research scope is the development of a supporting learning environment that would be able to monitor students' activity and identify low engagement and inefficient collaboration patterns, and assist students and instructors in improving the learning experience. The work presented in this paper reports on the first steps towards this goal.

As students walk through the phases and milestones of the project, they leave a very clear track of their involvement in the project. Analyzing the data acquired by usage logs of the software tools, reports sent by the students, and the consulting meetings students have throughout the assignment with TAs and instructors, we provide a framework for student assessment in software development projects. Our work on this framework breaks down to three research questions:
1. What can be accurately and efficiently monitored?
2. Which are the metrics related to the degree of students' individual engagement and the collaboration patterns occurred inside the groups?
3. How can these metrics trigger instructional interventions from the instructors or an intelligent tutoring system?

In the following, we present (a) the theoretical background of our approach, (b) the instructional method we applied, (c) the results on students' performance during the last two academic years, (d) discussion on the proposed framework, and (e) concluding remarks.

## II. BACKGROUND

From a theoretical point of view, our instructional approach draws from the project-based learning paradigm and engages students in collaborative activities in semester-long software development projects. Student monitoring is continuous but silent, meaning that although the data mirroring students' activity in the software tools used are available to students, it is not mandatory for the students to act upon them. Since our research scope is the development of a learning environment that would analyze and affect students' activity in real-time, our work is also related to intelligent tutoring systems.

### A. Collaborative Learning in Computer Science Education

Computer-Supported Collaborative Learning (CSCL) builds on the idea that peer interaction is a highly potent learning mechanism and can be realized and potentially enhanced by appropriately designed computer-based tools and scaffolds [18]. Research has shown that student interaction can indeed increase group performance and individual learning outcomes (e.g., [19],[20]). However, it has also been pointed out that productive learning interactions do not happen spontaneously within the team and research has consistently revealed that freely collaborating students may lack the competence to engage in fruitful learning interactions without external support and guidance [10]. To remedy this shortcoming, researchers have suggested various approaches to guide the collaboration activity, such as intelligent CSCL (e.g., [21]) and scripted collaboration, that is to provide learners with guidelines that specify, sequence, and assign roles and activities to small group of learners (e.g., [22], [23]).

Even in the case of scripted collaboration, one should expect to observe a range of collaboration patterns (e.g., [24]). One reason why this happens is that the typical collaborative task description always leaves room for students' self-organization of the activity at points where the task does not describe in detail what should happen. Even fine-grained and highly structured scripts leave some room for students' interpretation of an activity. This "distance" between the prescribed task and the actually implemented activity is highlighted in the literature by distinguishing between the ideal script (the activity as prescribed by the teacher), the mental script (the mental representation of the script that the group builds from teacher's prescription) and the actual script (the actual task and interactions that students engage into) [25]. It is worth mentioning that this task-activity distance effect is put in a more general perspective by authors who emphasize that the difference "is not a dysfunction of the learner rather than intrinsic to the notion of task and human activity" [26].

It would be impossible to define a comprehensive collaboration script engulfing every aspect of the software development process. To that extent, we fix approaches particular to the domain (e.g., agile development, types of artifacts like game-design-documents), but leave other aspects deliberately open. There are several methods in the software development world that are able to organize effectively the activity into distinct phases and support students in assuming different roles and reaching set goals. For example, pair programming (e.g., [27],[28],[29]) and collaborative quadruple programming [30] methods have been proposed for teaching software development.

As we present later, we applied a variant of the Scrum method for software development [34], allowing students a large extent of freedom in terms of self-organization and role assigning.

### B. Monitoring in CSCL

Monitoring students' activity in CSCL settings has been a widely implemented method in intelligent tutoring systems. Soller et al. [31] propose a model for the collaboration management cycle in their seminal work. The first phase in the cycle is the collection of the interaction data. The designer has to decide on the type of data collected, for example, whether they will be based on an activity-based analysis (requiring historical log), or on a state-based analysis [32]. In the second phase, a model of interaction is constructed, based on high-level variables and metrics that could efficiently represent the current state of interaction. Next, in the third phase, the current state of interaction can be compared to a desired model of interaction. The desired state needs to be defined by the designer using the variables of the current state. A desired state should be defined by variable values that would suggest meaningful interactions that would promote knowledge creation for the students. In the context of software development, an example of a desired state would include work distribution among the students and a regular pace of code commits. In the fourth phase, remedial actions are proposed, based on the differences observed between the current and the desired state of interaction. According to the implementation methods followed, corrective actions may be proposed by a human instructor or the system itself. Just before completing a full cycle and starting with a new iteration at the first phase, in the fifth phase of the cycle, the interaction assessment is evaluated and, if necessary, a new desired state is used until the learning goals are met. The re-evaluation can be done by a human tutor or the system, thus improving its own ability to diagnose student performance by directly analyzing student activity [33],[38].

Based on the feedback provided to the students during the activity, monitoring systems can be categorized into three groups:

Mirroring tools monitor students' activity and report that information directly back to them during the learning task through various formats (e.g., tables and graphical visualizations). The goal is to enhance students' self-awareness regarding their actions and behaviors. In mirroring tools, the human instructor is responsible for comparing the current and the desired model of interaction and deciding on an intervention.

In metacognitive tools, the information about the current state is presented along with information about the desired one. In this case, it is easier for the students to identify aspects of their interaction that may pose an issue and adapt their activity accordingly. As before, the students or a human instructor are responsible for evaluating and using this information to modify the learning activity.

Finally, guiding or coaching systems perform all the phases of the collaboration management process internally and are able to propose remedial actions directly to the students.

An example of a mirroring system in the context of software development would be a tool that presents to students a chart of their daily/weekly code commits. A metacognitive tool would also present a desired number of daily/weekly commits set by the instructor or based on the interaction model demonstrated by high-achieving students. Finally, a coaching system could alert the students via email and require a more detailed plan of daily/weekly commits. The analysis on the data we collected through the course will assist us in selecting appropriate metrics and reach meaningful conclusions regarding students' state.

III. METHOD

*A. Participants*

The study presents data for the last two terms of the course (Summer 2011 & Summer 2012). During this period, a total of 106 undergraduate students majoring in Computer Science in a three-year program participated initially in the activity (51 in the first and 55 in the second year). Students' participation was mandatory, since the activity was presented as a course on software development. In total, 6 students quit the course, while we excluded 13 more, because of their low attendance record in the course and/or their lack of engagement in the activity phases. This lowered the total population of students to 87 (43 for the first year and 44 for the second year).

In the beginning of the activity, we asked students to fill in an initial extensive, non-anonymous online questionnaire. Based on the answers from the questionnaire, 8 different scores are computed for each student, reflecting coding skill, domain knowledge, organizational skill, computer vision skill, sound editing skill, operating system affinity, negative motivation, and norm deviance. Students are then distributed by the instructors according to those scores into groups of 5 to 6 students. The distribution aims at minimizing inter-group variance for the first 5 scores, preferring the first two. The last three scores are used as tie-breaker. The scores are not disclosed to the students. Eventually, we formed 20 groups of 4-5 students (10 groups in 2011 and 10 in 2012).

*B. Design*

We performed exploratory analysis to identify appropriate variables to analyze students' patterns of individual engagement and collaboration. The treatment was the same for all the students and the dependent variables of the study were the quantitative metrics recorded by the software tools used, and the qualitative data recorded by TAs and instructors during their meetings and communication with the students.

*C. Procedure*

Table I presents the steps students have to go through during the activity. On the first week of the course, students attend a mandatory introduction lecture. In this lecture, students receive information about: the basic course layout, the topic and a set of high-level requirements for the software project they are going to implement, an overview of the used process model, and a list of tools that are used throughout the course. The topic of the course is the development of a computer video game for the domain of real-time strategy games. As nearly all students so far were unfamiliar with video game development, we could exploit consistently low base domain knowledge. Furthermore, video games are popular among students and as such it is easier for the students to imagine the workings of a finished product. This popularity also allows for a much higher motivation among the students.

At the end of the lecture, students are assigned a homework that should expose them to the basic technologies used during the lab and ensure that problems with tools are discovered early. The homework also entails reading some material about object-oriented programming, programming best practices, and video game design.

Once groups are assigned, group-mailing lists reaching all students in a group, their TA, and the instructors are generated. Then, the TAs contact with their students to agree on weekly meetings.

The lab uses a modified version of Scrum as process model. The weekly two-hour meetings with the TA serve as scrum meeting, where students define their goals that are to be achieved within one week in a Sprint backlog and discuss if the goals for the previous week have been achieved. TAs ensure that work is distributed evenly, that important tasks are not forgotten, and help with technical and organizational questions.

TABLE I. ROADMAP FOR THE SOFTWARE LAB COURSE

| Week | Activity | Design | Impl. |
|------|----------|--------|-------|
| 1 | Fill out questionnaire | | |
| 2 | Submit homework | x | |
| 3 | Submit game design document (beta) | x | x |
| 4 | Presentation #1: Game idea | x | x |
| 5 | Submit class diagram (beta) | x | x |
| 5 | Milestone #1: Basic prototype | x | x |
| 7 | Lecture "Clean Code" | x | x |
| 7 | Milestone #2: Interactive prototype | x | x |
| 8 | Submit game design document (final) | x | x |
| 8 | Code review #1 | x | x |
| 9 | Presentation #2: Software (beta) | | x |
| 9 | Submit software (beta) | | x |
| 10 | Submit class diagram (final) | | x |
| 10 | Code review #2 | | x |
| 10 | Milestone #3: Beta version | | x |
| 13 | Milestone #4: Final product | | x |
| 14 | Presentation: Software (final) | | x |
| 14 | Submit software (final) | | x |

After the first week, the course is structured in two overlapping phases, design and implementation. The design phase lasts from week 2 to week 8, the implementation phase from week 3 to week 14.

In the design phase, each group creates an artifact defining their own set of requirements that have to satisfy the given high-level requirements. This artifact is a domain-specific software requirement specification document called game design document (GDD). Each group also designs a software architecture that allows the implementation of their design. Here, only class diagrams are used as artifacts.

In the implementation phase, the students follow a domain-specific project plan with five milestones and implement their project accordingly. In this phase, the instructors also conduct two 2-hour code reviews with each group, explaining the use of certain metrics to discover possible errors and possibilities for improvement in the code base of the groups. During the code reviews, students are exposed to metrics generated by NDepend (www.ndepend.com) and Sonar (i.e., LCOM4, afferent and efferent coupling, cyclomatic complexity, and package stability) and statistics generated by StatSVN (www.statsvn.org) (i.e., commit patterns, ratio between added and modified content per user, total code contribution per user). The implementation phase ends with the final submission of a deployable version of the software.

We also include three mandatory presentation dates during the course. There, the groups present their current state publicly to their fellow students, the TAs, and the instructors. The audience is also allowed to ask questions.

Lastly, each student writes a short report 1-3 times a week presenting: the tasks he is currently dealing with, the issues he faces, the time estimate for completing each task, and how much time he has already invested in each task. The reports are sent to the group mailing list. Weekly evaluations of the group by the instructors, based on a traffic metaphor, determine how many reports each member of the group has to write (green: 1, yellow: 2, red: 3).

Lastly, each week students are offered a 4-hour session of supervised programming in the computer lab of the university. During these sessions, instructors and TAs are present in the lab and can help solve technical problems more hands-on.

In order to complete the course successfully, the groups have to deliver a complete, bug-free version of their video game. We measure completeness by comparing the final product with the groups' GDD, and bug-freeness through multiple playthroughs. Admission to grading is managed on a per-student basis. Students have to send reports, attend group meetings and the presentations, and contribute continuously to the groups' project. The last part is shown through man-hours in reports, commit patterns in the repository, and the content of commits. Every admission criteria is evaluated weekly.

TAs serve as monitors for their groups, e.g., they inform the instructors of possible problems with single students or whole groups in a weekly 2-hour meeting. Because those problems are always group-specific, instructors need to continuously monitor the current group situation, and prepare specific interventions if problems surface.

After the end of the course, students can optionally fill in the final questionnaire, stating their opinions about different aspects of the whole activity.

*D. Tools and Instruments*

We designed the initial questionnaire to better understand the students' backgrounds and strengths. It is comprised by 8 categories: personal data, programming experience, software development experience, domain knowledge, practical experience, motivation, and questionnaire quality. The instrument included both closed (Likert scale -2 to 2) and opened-type questions.

*Personal data* section refers to biographical data, such as age, year of study, major, and email address. *Programming experience* section refers to students' experience with various programming languages, and operating systems. *Software development experience* section is about familiarity with different software development tools, self-assessment of understanding of object-oriented programming, and experience with different aspects of software quality assurance. *Domain knowledge* section contains questions about students' engagement with video games, the preferred gaming platforms, and the preferred game genre. *Practical experience* section focuses on students' experiences in industry, such as roles assumed in real-world projects (e.g. manager, developer, tester), paid work, type of client (e.g., university, company), and time invested in projects. *Motivation section* is about the reasons for signing up for the course. Finally, *Questionnaire quality* section refers to the overall quality of the questionnaire.

In the last third of the course, students can complete an anonymous online evaluation questionnaire used for the official evaluation of the university, hence only a part of it is designed for the course and this work. Those items were designed to evaluate the performance of the TAs and the instructors, the performance and suitability of the various tools provided, and the general level of acceptance regarding the different aspects of the software lab such as the overall workload, the amount of support provided, and the impact of the course to students.

Besides the various information sources, students are given multiple software tools that support their project. During the course, students have access to a Wiki, where information regarding game development, programming language, tool usage and course organization is available to them. Furthermore, they are continuously encouraged to ask their group members, their TAs, or the instructors if they face problems or have questions. To this end, we set up a group-chat software (IRC) where instructors and TAs can be asked anonymously, outside of office hours, and on weekends.

Every group has access to a version-control (Subversion: subversion.apache.org) and bug-tracking system (Trac: trac.edgewall.org). A build server generates nightly builds of the repositories, runs unit tests, and informs the students via email about issues with their project. On the client side, students use Microsoft Visual Studio 2010 as IDE (www.microsoft.com/visualstudio/eng/downloads#d-2010-express). One of the technical requirements is the usage of C# as programming language together with the game development framework XNA 4.0 (www.microsoft.com/en-us/download/details.aspx?id=23714), both for .NET 4.0. Additionally, we use JetBrains ReSharper (www.jetbrains.com/resharper) as plugin to the IDE. It allows the on-the-fly enforcement of coding conventions and also serves as a kind of tutoring system, as it uses heuristics to suggest language usage opportunities to the students. Besides those mandatory tools, students have access to web-based analysis software that informs them about various metrics and statistics regarding their source code. StatSVN generates daily per-user statistics from the repositories and compares them among the developers. Codehaus Sonar (www.sonarsource.org ) generates daily source code metrics (like lines of code, cyclomatic complexity of methods or types, etc.) from the nightly builds. Doxygen (www.stack.nl/~dimitri/doxygen) generates daily source code documentation together with dependency diagrams from comments in the code.

*E. Data Analysis*

In the analysis, we collected data from the students, the TAs, and the software tools used in the activity. Students' data came through the weekly reports they had to submit and the emails they sent to the mailing list. Students informed us about the list of tasks they performed each week and the time they allocated to each task. TAs were able to give us more qualitative information, because of the frequent communication they had with students. So, opinions, unresolved issues, and any kind of problems were discussed with them. Perhaps the more valuable input regarding TAs' role in the activity was to assist the instructor in determining the status of a group. For this, a traffic light metaphor was adopted. Green status means that the instructors were confident that this group would reach the next milestone without any trouble, yellow status means there is room for improvement and the instructors are unsure if the group reaches the next milestone, and red status means that the instructors are sure the group will not reach the next milestone and/or has not reached the last one.

The bulk volume of our analysis though, came from the log files of the software tools used. For data analysis, we collected weekly information from software repositories such as Subversion and Trac (Table II). All data are collected on a weekly basis to observe the change of collaboration patterns over time. A commit in Subversion shows the progress history of a project. By simply counting the number of commits, we can estimate the progress made by each group or student. We count the number of commits in different levels (group or student) and targets (commits for source code, binary, or both). In this way, we can analyze how much each group or student contributes to which tasks of a project.

The use history of Trac can provide the collaboration data of students to deal with reported issues and Wiki pages in terms of project management. To analyze the use pattern of Trac, we count the POST logs from Trac http logs. The POST logs represent the number of added or edited posts such as an issue and a Wiki page. We also computed Trac access counts in both group and student levels.

To analyze the frequency of communication, we count the number of emails sent in each group on a weekly basis. With this data, we can investigate whether communication frequencies can affect grades or not.

In addition, we consider the number of files, which are modified by the same number of students. We regard the students who commit a modified file in Subversion as owners of the file. For example, if only the files A and B are individually modified and committed by four different students in a certain week, the number of files by four owners is 2 in that week. By analyzing this data, we can investigate students' collaboration patterns on project files; that is whether students tend to work on same files together or not over weeks.

TABLE II. TYPES OF WEEKLY DATA COLLECTED

| Type | Target | Description | Level |
|---|---|---|---|
| # of commits | Source code (.cs) | Weekly counts of commits of all, source code, or binary files. | Group, Student |
| | Binary (.jpg, .png, .bmp, and .wav) | | |
| # of Trac access | POST | Weekly Trac access counts collected from http logs. | Group, Student |
| # of emails | All sent emails | Weekly count of all emails sent by each group | Group |
| # of files by owners | Source code (.cs) | Weekly count of the number of files owned by N students | Group |
| | Binary (.jpg, .png, .bmp, and .wav) | | |

## IV. RESULTS

### A. Student Profile and Performance

Results analysis on students' answers in the initial questionnaire revealed that students had programming experience in C++ (80.18%), since they were already in their fourth semester of studies. However, a much lower number of students were familiar with C# (16.98%). As such, the technical requirements posed by the course were highly demanding for most of the students. Additionally, students had mixed opinions about their programming skills ($M = -0.11$, $SD = 0.96$). Finally, student motives for participating in the course were in descending order the selected topic (62.26%), the chance to work collaboratively with other (53.77%), and to learn programming (50.94%).

All students in a group received the same grades. Despite the demanding nature of the course, students' performance was satisfying ($M = 7.50$, $SD = 2.52$), with 6 groups receiving full grade and only 3 borderline passes.

### B. Individual Engagement

Based on (a) the data gathered through information retrieval techniques on the software tools used during software development, (b) students' self-reported information in weekly reports, and (c) discussions between students and TAs in the consultation meetings, we analyzed students' activity in two levels. Our first concern was to evaluate how each individual was engaged into the activity. In other words, what kind of roles and responsibilities did each group member acquired, how much effort each one contributed to the final product, and which were the dominant working patterns that emerged? On a second level, we focused on the collaboration patterns occurred inside the groups, meaning the distribution of workload, the amount of communication, and the product developed throughout the 14 weeks of the activity.

Starting with the roles each student in a group acquired, we based our analysis first on the number of commits the students performed, then on tickets issued, and lastly on the self-reported number of hours for work on the project (Table III).

As expected, the source files committed were significantly more than the binary files. Binary files are usually images related with the user interface. As such, work on binary files is expected to be less. While all students committed code, an interesting pattern appeared in some of the groups. In 12 groups, the role of the designer was played almost exclusively by 1 or sometimes 2 students who were responsible for more than the 80% of the visual work done. The distinct distribution of workload did not affect groups' performance as this way of work was evident in both high and low achieving groups. In total, 17 students took the role of a lead designer, while in the rest 8 groups this responsibility was equally assumed by the members.

TABLE III. METRICS OF INDIVIDUAL ENGAGEMENT

|  | Per Student (n = 95) | | Per Group (n = 20) | |
| --- | --- | --- | --- | --- |
|  | *M* | *(SD)* | *M* | *(SD)* |
| **Code** | 113.60 | (86.50) | 590.95 | (150.24) |
| **Binary** | 10.08 | (10.32) | 51.90 | (24.37) |
| **Tickets** | 104.18 | (118.45) | 494.85 | (349.57) |
| **Hours** | 163.09 | (71.36) | 775.89 | (196.21) |

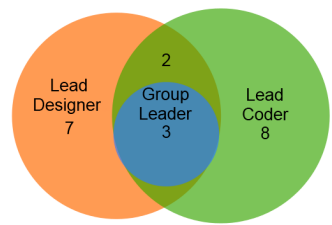

Fig. 1. Group distribution according to different roles

Similarly, we can identify role of a lead coder, the student who was responsible for almost half of the code committed by his group. Hence, based on the number of code commits, the role of lead coder appeared in 13 groups, while the workload for programming was balanced in the remaining 7 groups. Once again, the distinction between groups based on the role of a lead coder is not correlated to the groups' performance and or the appearance of lead designers. Eventually, there were only 4 groups where both roles of lead designer and coder appeared and in 3 of them, both roles were played by the same students. We refer to these students as group leaders since they are responsible for most of the work done in their teams. Figure 1 presents the distribution of the groups based on the existence of lead designers, lead coders, and group leaders.

The appearance of lead roles allowed other students to stay behind. We borrow the term "free riders" from [35] to refer to students that applied low engagement strategies. These students are usually responsible for less than the 10% of all the work done in the project, while in some cases the percentage of commits done by these students was even lower than 3%. Free riders were almost a universal phenomenon, as there were 15 groups with 19 free riders in total.

Two more metrics for estimating students' engagement were the number of tickets posted during the activity and the working hours students reported in their weekly reports. Issuing tickets suggests that students have spent time on the project to be able to identify specific tasks that needed to be carried out. In that sense, our assumption is that a high number of tickets could suggest a deeper understanding of the developing project.

Commits and tickets were easy to count, by looking into the log data of the software tools. However, for the actual amount of time spent on the project, we had to ask student. Although we cannot directly link the amount of time spent with the quality of the outcome, we accept the self-reported metric of hours spent on the project as an estimate for students' engagement.

A descriptive level analysis indicated that the total numbers of commits, tickets, and hours for each student may be correlated. Indeed, statistical analysis revealed that code

TABLE IV. PEARSON'S CORRELATION COEFFICIENT

|         | Code | Binary | Tickets | Hours   |
|---------|------|--------|---------|---------|
| Code    | 1    | 0.210* | 0.334** | 0.471** |
| Binary  |      | 1      | 0.048   | 0.246*  |
| Tickets |      |        | 1       | 0.233*  |
| Hours   |      |        |         | 1       |

* p<0.05, ** p<0.01

commits, binary commits, tickets, and hours are all significantly correlated (Table IV), with the only exception being a lack of correlation between binary commits and tickets, which is normal, since most tickets refer to coding issues rather than design. The correlation between the commits and the self-reported hours, suggests that students answers in the reports were true, since the self-reported working pattern seems to mirror what is monitored through the software log.

Figure 2 presents how students distributed their workload throughout the period of the 14 weeks. Usually, work on visual parts starts earlier, but, after week 4, work on code and design goes almost in parallel. There is continuous rise and then a significant drop during week 8. This is when students have to submit the final game design document, and, as expected, their efforts during this week are allocated mostly into finalizing this document. Also in week 8 students go through their first code review session with their TA, and this is where they get a deeper analysis of issues regarding their produced code. The degree of effort increases during the next couple of weeks, until the second code review in week 10. Between weeks 10 and 13, students' progress is still high but with a considerably decreased pace. This could be attributed to the lack of any major deadline. On week 13, they discuss with their TA the state of the progress of the project and issues for the upcoming presentation. After that there is a final sprint towards week 14, where they finalize everything and get ready to submit their project.

*C. Collaboration Patterns*

To understand the collaboration patterns that emerged in the groups and also to identify links between groups with similar performances, we took into account - on top of code and binary commits, tickets, and hours - three additional variables for each group for each week: (a) group state according to the traffic light metaphor, (b) number of emails sent in the mailing list of the group, and (c) the percentage of files with shared ownership. As we mentioned earlier in the data analysis, shared ownership suggest work overlapping and possible miscommunication among the group members. Also, a sudden peak on the ownership metric could suggest a rapid re-allocation of responsibilities, indicating another type of troubled collaboration.

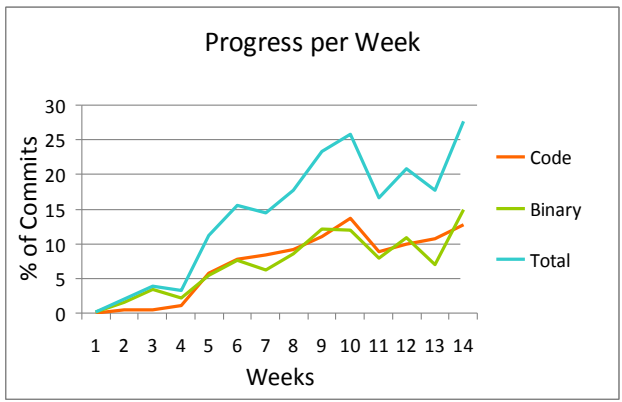

Fig. 3. Effort distribution throughout the course

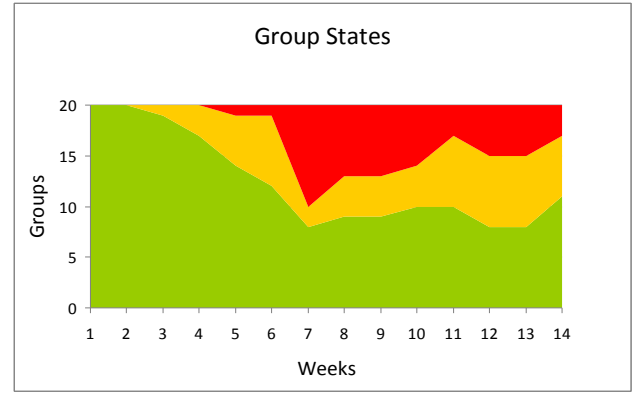

Fig. 2. Group States during the Course

Figure 3 presents how groups' states changed during the 14-week period. As we mentioned earlier, the state is based on the TA's estimation about the probability that a certain group will be able to reach its goals. It is obvious that right before the first code review session and the submission of the final game design document, half of the groups went into red. Most of them were able to recover during the following weeks.

Another interesting case was a group whose state became red after week 4. This group was not able to recover and was trailing behind through the whole assignment, having also the lowest grade in the class. Statistical analysis showed that the last state of each group was strongly correlated with the final grade of the group in the course (r = 0.696, p = 0.001). Case by case analysis also revealed that the performance of groups that reported lower working hours in the project than others by week 6 and were also behind in terms of commit numbers, tend to be negatively affected in terms of group state and final course grade.

While tickets were used to manage project tasks, the group mailing list served the purpose of overall communication, task allocation, scheduling meetings, QA sessions, and so on. The number of emails sent per group increased during the first few weeks and remained stable throughout the course (M = 380.65, SD = 205.91). According to TA's reports, there were also groups where the group mailing list was used interchangeably with the ticketing tool, while in others the issue of a new ticket could trigger more emails. We tested this finding for the latter by calculating the cross-correlation coefficient between the time series of tickets and emails each week and results showed that the two variables were significantly correlated (p<0.05).

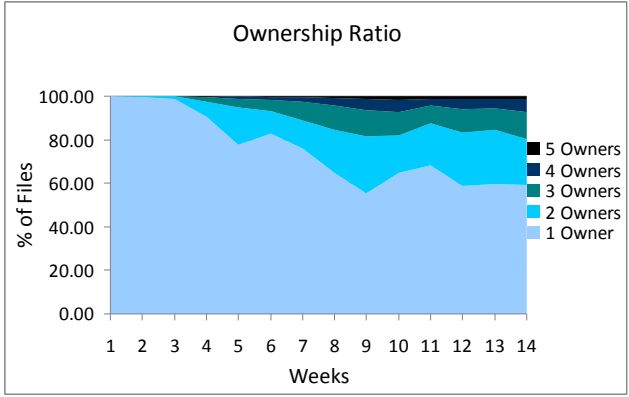

Fig. 4. Group States during the Course

Finally, we analyzed groups' collaboration patterns in terms of ownership. As expected, the percentage of files that were modified by more than one group members during the same week grew throughout the 14-week period (Fig. 4). After examining the emerging lead roles in the groups, our initial hypothesis was groups with and without a leader would have different ownership patterns. However, cross-correlation coefficient analysis on the time-series of these two sub-groups revealed that they were almost identical (p<0.05). Additionally, in 9 groups, the patterns of code commits and ownership ratio were significantly correlated, suggesting that the more the student committed, the more they started to modify each other's files.

*D. Student Opinions on the Course*

Table V presents selected results from the final questionnaire. For both years, more than half of the students participated in the questionnaire.

According to students, the workload was too high in relation to the credit points for the course. This result is expected, as other courses in the curriculum give an equal amount of credit, where a final exam is the only assessment. However, when asked about the relation of workload to learning outcome, students tend to be satisfied with the workload. Another interesting result is that students assess their contribution to the project as valuable for the team. This indicates that students see themselves as a valuable part of the team and that their collaboration with others had a positive effect on their work.

Most of the provided tools were positively received by the students. In particular, SVN and ReSharper are considered highly valuable. The large acceptance of ReSharper indicates that we can put more emphasis on code quality and supporting tools in the future.

The IRC channel was also very popular among few students, however, it was not widely used. We are thinking about moving more face-to-face tutoring towards the use of IRC to encourage students to use this tool.

TABLE V. SELECTED RESULTS FROM THE FINAL QUESTIONNAIRE

|  | 2011 | | | 2012 | | |
| --- | --- | --- | --- | --- | --- | --- |
|  | *M* | *(SD)* | *n* | *M* | *(SD)* | *n* |
| **Workload in relation to credits** | 1.88 | (0.31) | 35 | 1.44 | (0.56) | 27 |
| **Workload in relation to learning outcome** | 0.60 | (1.04) | 35 | 0.42 | (0.88) | 26 |
| **Value of my contribution for the team** | 1.00 | (0.93) | 34 | 1.33 | (0.60) | 27 |
| **Usefulness of provided tools** | | | | | | |
| **ReSharper** | 1.54 | (0.60) | 33 | 1.30 | (0.95) | 20 |
| **Trac** | 0.42 | (1.14) | 28 | 0.12 | (1.20) | 24 |
| **SVN** | 1.74 | (0.43) | 27 | 1.47 | (0.75) | 19 |
| **IRC Channel** | 1.61 | (0.62) | 13 | 1.14 | (0.83) | 7 |
| **Lab computers** | 1.11 | (1.09) | 18 | 1.30 | (0.95) | 20 |

Lab computers considered useful as well by the students. This is, in a way unexpected, as most of the students have more powerful equipment and laptop computers. However, this public space with the necessary equipment to work collaboratively on the project was widely used by the students.

V. DISCUSSION

Regarding our first research question on what can be accurately and efficiently monitored, our work presents a significant amount of data. Starting from the tools used by the students, metrics such as the number of commits, tickets, and emails, and the ratio of shared ownership could collectively paint a picture of the engagement of individuals in the activity. Although one can argue that volume and submission patterns are not necessarily related to the quality of the work produced, a low number of commit and an irregular pattern of submission should catch attract trigger instructor's attention. Information retrieval techniques can go even deeper.

Throughout this analysis we identified additional sources of data, which have not been used for this paper. Our results show, for example, that the status of groups assigned by their TAs can be valuable to predict students' performance. However, the weekly reports students send according to their traffic light state, are only used for the self-reported working hours. Information about which tasks have been reported as complete or incomplete has not been considered, as the template for the report did not allow us to mine this data automatically. A certain future improvement would be to replacing the emailed reports with web-based forms, and to define a standardized encoding for the tasks in such a project to get more information about the individual progress of each student.

The ticketing system turned out to be less useful than expected, as results showed that most of the ticketing activity took place even before the students started programming. As ticketing is something that is widely used in the professional world, we need to put more emphasis on encouraging Trac usage from students. Here, the aforementioned standardized encoding of tasks could also prove useful.

Regarding the metrics related to patterns of individual engagement and collaboration, we suggest that the number of commits, the self-reported working hours, and the posted tickets could provide first level evidence on individual participation in a group. Based on these metrics it was easy for us to identify the emergence of lead roles and free rider behaviors.

The combination of the above with group metrics such as the group state, the number of emails, and the file ownership ratio could provide insights on the collaboration patterns. The three groups with the lowest grades reported less working hours than average during the first weeks of the coding phase. Furthermore, all these teams started to commit to their SVN relatively late, compared to other groups. Additionally, while it is somewhat expected that students are going to divide the workload and assume distinct roles choosing specialties, the appearance of a group leader could suggest problematic allocation of workload, since one student is responsible for most of the work, and this allows others to stay behind. Finally, the information about ownership gave us indications about students' collaboration, but it is clear that more detailed information is necessary in order to benefit significantly from this metric. Future research can focus on information about the shared ownership not only in general in the group level, but also between group members. In this way, we will be able to identify more patterns of peer interaction within the group.

Finally, regarding instructional interventions, our data suggests that code reviews and programming sessions turned out to be an effective instrument to increase the overall activity of a group. We observed that regular deadlines, such as prototype presentations, increase the activity of students. This is in line with the idea of having an executable prototype after each sprint in the Scrum process. We will use this to improve our variation of the Scrum process for this lab and develop better assessment techniques for the outcome of sprints.

## VI. CONCLUSIONS AND FUTURE WORK

The primary goal of the course is to introduce students to collaborative software engineering. The students were exposed to a variety of tools that are commonly used in professional software engineering. By using these tools, students generated a detailed trace about their individual engagement, which we use to search interaction patterns that allow us to predict problems in the collaboration behavior within their group.

From the obtained data we are able to identify common problems such as the free rider phenomenon. Several other issues of individual groups regarding bad collaboration behavior could be monitored along the collected data. However, in several cases, the data is not enough to effectively identify individual or collaborative patterns that can be useful for future iterations of the course. Here, we identify several shortcomings in our course design, which could be mainly attributed to the high degree of freedom allowed to students. First, the data obtained from the Trac system was below our expectation. We will adjust the course design towards a more mandatory use of Trac. Second, the weekly report emails could not be analyzed automatically due to a lack of structure. Even though we could extract the individual working hours reported, we could not identify which tasks have been completed in a week. Here, we propose a shared taxonomy between Trac, report emails, and the requirements in the GDD. Such a link will provide very detailed information about the student's individual progress. Third, the ownership data did not give the expected results. In the future, we will compute ownership between groups of students to find patterns of subgroups.

Overall, we are satisfied with the outcome, as the data analysis on students' activities are in line and can predict to an extent the grades we assigned to each group. Equally important though is the fact that students seem to appreciate the instructional method and the learning benefits of our approach.

ACKNOWLEDGMENT

Supported in part by Macau Science and Technology Development Fund in the context of (a) the COLAB project, and (b) the PPAeL project, File No. 019/2011/A1.